\documentclass[12pt,a4paper]{iopart}

\usepackage[utf8]{inputenc}
\usepackage{iopams}
\usepackage{graphicx}
\usepackage{microtype}
\usepackage{xcolor}
\usepackage[numbers,square,comma,sort&compress]{natbib}

\definecolor{hl}{rgb}{0.106,0.604,0.569}
\usepackage[
    colorlinks=true
   ,citecolor=hl
   ,linkcolor=hl
   ,urlcolor=hl
   ,pdfborder={0 0 0}
   ,pdfusetitle
]{hyperref}

\newcommand{\text}[1]{\mathrm{#1}}
\newcommand{\eqref}[1]{(\ref{#1})}

\begin{document}

\title[Ferroelectric quantum phase transition with cold polar molecules]{Ferroelectric quantum phase transition with cold polar molecules}

\author{Markus Klinsmann, David Peter and Hans Peter Büchler}
\address{Institute for Theoretical Physics III, University of Stuttgart, Germany}

\ead{buechler@theo3.physik.uni-stuttgart.de}

\date{\today}

% 67.85.-d - Ultracold gases, trapped gases
% 37.10.Jk - Atoms in optical lattices
% 05.30.Jp - Boson systems
% 77.80.-e - Ferroelectricity and antiferroelectricity

\begin{abstract}
We analyze a system of polar molecules in a one-dimensional optical
lattice. By controlling the internal structure of the polar molecules with
static electric and microwave fields, we demonstrate the appearance of
a quantum phase transition into a ferroelectric phase via spontaneous
breaking of a $U(1)$ symmetry. The phase diagram is first analyzed
within mean-field theory, while in a second step the results are
verified by a mapping onto the Bose-Hubbard model for hard-core
bosons. The latter is studied within the well-established bosonization
procedure. We find that the ferroelectric phase is characterized by
(quasi) long-range order for the electric dipole moments.
\end{abstract}

\maketitle

The experimental realization of cold atomic and molecular gases with strong dipole-dipole interactions has
opened up new avenues for the exploration of strongly correlated states of quantum matter \cite{Lahaye2009,Ni2008b}.  A characteristic   property
of dipolar interactions is their  anisotropic behavior as well as the slow decay in space giving rise to ordering and formation of
large scale structure. Of special interest would be the spontaneous ordering of electric dipole moments giving rise to a ferroelectric state of matter.
Here, we propose a setup with cold polar molecules which gives rise to such a ferroelectric state of matter, characterized by a macroscopic dipole moment. Such a state is analogous to the well-established phenomenon of ferromagnetism appearing for materials with magnetic moments.

Important properties of cold polar molecules are the existence of a permanent  electric dipole moment $d$ and the presence of  the rotational energy splitting $B$. It is this rich internal structure in combination with the high control  via static electric fields and microwaves,
which form the basis for the design of many strongly correlated states of matter \cite{Buchler2007,Bonnes2010,Buchler2007a,Gorshkov2011c,Capogrosso-Sansone2010,Burnell2009,DallaTorre2006a,Lin2010a}.
On the other hand, it is also the rotational splitting which  prevents the spontaneous ordering of the electric dipole moments for characteristic energy scales  $d^2/a^3 < B$ where $a$ denotes the inter-particle distance.  This condition is naturally satisfied for cold polar molecules, and its violation would require inter-particle distances of a few nanometers.  However, the rotational splitting can be effectively reduced by a weak microwave field, coupling the different
rotational levels. A similar procedure has previously been proposed for a Bose-Einstein condensate of polar molecules \cite{Lin2010a}.  Then, the ferroelectric phase corresponds to a coherent superposition of two rotational states driven by the dipole-dipole interaction. It gives rise to a macroscopic dipole moment oscillating in time with the frequency of the weak microwave field.

In this manuscript, we study the appearance of the ferroelectric phase for polar molecules within a one-dimensional optical lattice.  A static electric field is imposed to induce a permanent dipole moment perpendicular to the lattice. A circularly polarized microwave field reduces the energy of the relevant internal levels of the molecules. The phase diagram is first analyzed within mean-field theory, while in a second step the results are verified by a mapping onto the Bose-Hubbard model for hard-core bosons. The latter is studied within the well-established bosonization procedure. We find a quantum phase transition leading to a ferroelectric phase via spontaneous breaking of the $U(1)$ symmetry. The low energy excitations exhibit a linear dispersion relation providing (quasi) long-range order at zero temperature.

\begin{figure}[t]
    \centering
    \includegraphics[width=0.99\textwidth]{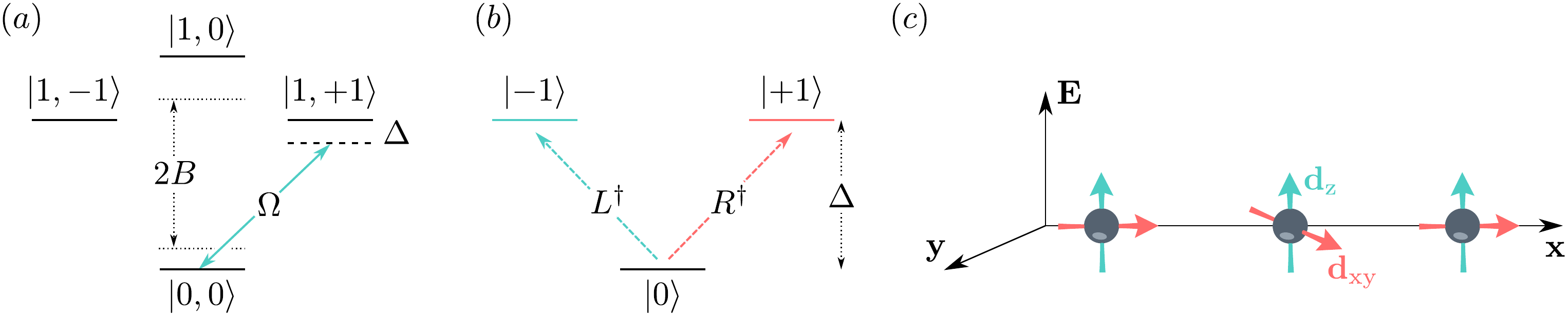}
    \caption{
    (a) Internal level structure of each molecule with the rotational
    states $|J,m\rangle$ and the microwave coupling with Rabi frequency
    $\Omega$ and detuning $\Delta$.
    (b) Relevant states in the rotating frame. The energy separation is
    effectively reduced to $\Delta$. Action of the operators $L^\dag,
    R^\dag$ is indicated.
    (c) Setup: polar molecules are confined in a 1D optical lattice with
    one molecule per site. An external electric field $\mathbf{E}$ in
    $z$~direction induces finite dipole moments $\mathbf{d}_z$. Within
    the ferroelectric phase, the molecules acquire a finite, oscillating
    dipole moment $\mathbf{d}_{xy} \sim \mathbf{e}_x \cos(\omega t)$
    along the $x$~direction, where $\omega$ is the frequency of the
    microwave.
    }
    \label{fig1}
\end{figure}

We start with the description of the setup. The polar molecules are confined in one dimension
along the $x$-axis,  and trapped by an optical lattice with one polar molecule on each lattice site, see FIG.~\ref{fig1}. The relevant internal structure of each polar molecule is given by the rotational degree of freedom and  the Hamiltonian for a single polar molecule reduces to a rigid rotor for the dipole moment
\begin{eqnarray}
    H_{\text{rot}}=B \mathbf{J}^2 - \mathbf{d}\cdot\mathbf{E}, \label{eq:Hrot}
\end{eqnarray}
with the angular momentum operator $\mathbf{J}$ as well as the rotational constant  $B$. In addition, the second term accounts for the coupling of the dipole moment $\mathbf{d}$ to the electric field $\mathbf{E}$. For vanishing electric field $\mathbf{E}=0$, the eigenbasis of $H_{\text{rot}}$ is formed by the
%spherical harmonics $Y_{Jm}$
states $|J,m\rangle$,
with the quantum number $J$ for the total angular momentum and $m$ its projection along the $z$-axis, i.e., $J_{z} |J,m\rangle= m |J,m\rangle$. It is important to stress that  these states do not exhibit a dipole moment, i.e.
$
\langle J,m |\mathbf{d}| J,m \rangle=0,
$
with $\mathbf{d}$ denoting the dipole operator.  In the following, we are interested in a setup with a finite static electric field along the z-axis, which leads to a mixing of states with different quantum numbers $J$. In particular, this splits the threefold degeneracy of the $J=1$ manifold. The numerical determination of the new exact eigenstates is straightforward. Next, we restrict the analysis to the three states with the lowest energy, which can be characterized by the quantum number $|m\rangle$, i.e., the ground state with $m=0$, and two degenerate excited states with $m=\pm 1$, see FIG.~\ref{fig1}.

It is important to stress that the energy separation between the ground state and the excited states is of the order $B$,  which is very large compared to the characteristic energy scales of the polar molecules given by the interaction. Therefore, we reduce this energy by applying a weak circularly polarized microwave field coupling the ground and excited states. This coupling is only required to allow for relaxation and all results are presented in the regime where this coupling is adiabatically turned off.
Within the rotating frame, measuring energies with respect to the ground state, the Hamiltonian projected onto the relevant states reduces to $H = -\Delta \, P$ with the projection $P=|{-1}\rangle\langle{-1}|+|1\rangle\langle1| $ onto the degenerate manifold, and $ \Delta=\Delta(\omega,E^{\text{dc}})$ the energy difference between the ground state and the rotationally excited states. The latter  depends on the frequency $\omega$ of the microwave field, and the static electric field $E^{\text{dc}}$. Note that the microwave field allows us to tune $\Delta$ to arbitrary values and especially also into the regime $\Delta > 0$ with the degenerate manifold $m=\pm 1$ exhibiting the lower energy.

The interaction between the polar molecules residing on each site of the one-dimensional lattice is well accounted for by the dipole-dipole interaction
\begin{eqnarray}
V_{i j}
=\frac{\mathbf{d}_i\cdot\mathbf{d}_j}{|\mathbf{r}_{ij}|^3} - \frac{3\left(\mathbf{d}_i  \cdot\mathbf{r}_{i j}\right)\left(\mathbf{r}_{i j}\cdot\mathbf{d}_j\right)}{|\mathbf{r}_{ij}|^5}, \label{eq:Vdd1}
\end{eqnarray}
%
%
%%
%\begin{eqnarray}
%V^{dd}_{ij}
%=\frac{\mathbf{d}_i\cdot\mathbf{d}_j-3\left(\mathbf{d}_i\cdot\mathbf{e}_R\right)\left(\mathbf{e}_R\cdot\mathbf{d}_j\right)}{|\mathbf{R}_{ij}|^3}, \label{eq:Vdd1}
%\end{eqnarray}
%
with the relative position $\mathbf{r}_{i j} = \mathbf{r}_{i} -\mathbf{r}_{j}$ between the two polar molecules at site $i$ and $j$.
The projection of the full dipole-dipole interaction onto the relevant internal states $|m\rangle_{i}$ provides the full many-body Hamiltonian of our system. This projection is most conveniently achieved by introducing the operators
$R^{\dag}_{i}=|1\rangle\langle 0|_{i}$ and $L^{\dag}_{i}=|{-1}\rangle\langle 0|_{i}$, as well as the projectors
$P_{i}=|{-1}\rangle\langle{-1}|_{i}+|1\rangle\langle1|_{i} $ and $Q_{i} = |0\rangle \langle 0|_{i}$, see FIG.~\ref{fig1}\,(b).
Within the rotating frame and applying the rotating wave approximation, the projection of the dipole-dipole interaction reduces to
\begin{eqnarray}
    V_{ij} =  \frac{1}{|\mathbf{r}_{ij}|^{3}}
    &\Big[ \big( d_{11} P_{i} + d_{00} Q_{i} \big) \big(d_{11} P_{j} + d_{00} Q_{j}\big)  \\
 &  \quad -\frac{d_{10}^2}{2} \big(  L_{i}^{\dag} (L^{\vphantom\dag}_{j} - 3 R^{\vphantom\dag}_{j}) + R_{i}^{\dag} (R^{\vphantom\dag}_{j} - 3 L^{\vphantom\dag}_{j}) + \text{h.c.} \big) \Big]
\nonumber
 \end{eqnarray}
with the dipole matrix elements $d_{m m'}=|\langle m| \mathbf{d}|m'\rangle|$. The terms in the first line account for the static dipole-dipole interaction between the different polar molecules due to the induced dipole moments, while the terms in the second line describe the resonant exchange processes.  Then, the full many-body Hamiltonian reduces to
\begin{eqnarray}
    H= -\Delta \sum_i P_{i} + \sum_{ i< j}V_{ij}, \label{eq:Htotal}
\end{eqnarray}
which has a global $U(1)$ symmetry for remaining invariant under a multiplication of the excitation operators $R, L$ by a phase factor.

We summarize the qualitative behavior of the system in the following: The crucial parameter of this discussion is the detuning $\Delta$. Depending on its value we divide the phase diagram of the system into three different regimes. For large negative values of $\Delta$ an excitation is too costly and all the molecules stay in the ground state~(I). Increasing $\Delta$ we arrive at a point where it is energetically favorable to excite some molecules, resulting in a finite in-plane polarization~(II). The number of excited molecules then slowly grows until it reaches unity at a certain value of $\Delta$. At this point, the strength of the induced field has dropped from a certain maximum value back to zero. A further increase in $\Delta$ finally leaves the system unaffected~(III).

\begin{figure}[t]
    \centering
    \includegraphics[width=0.6\textwidth]{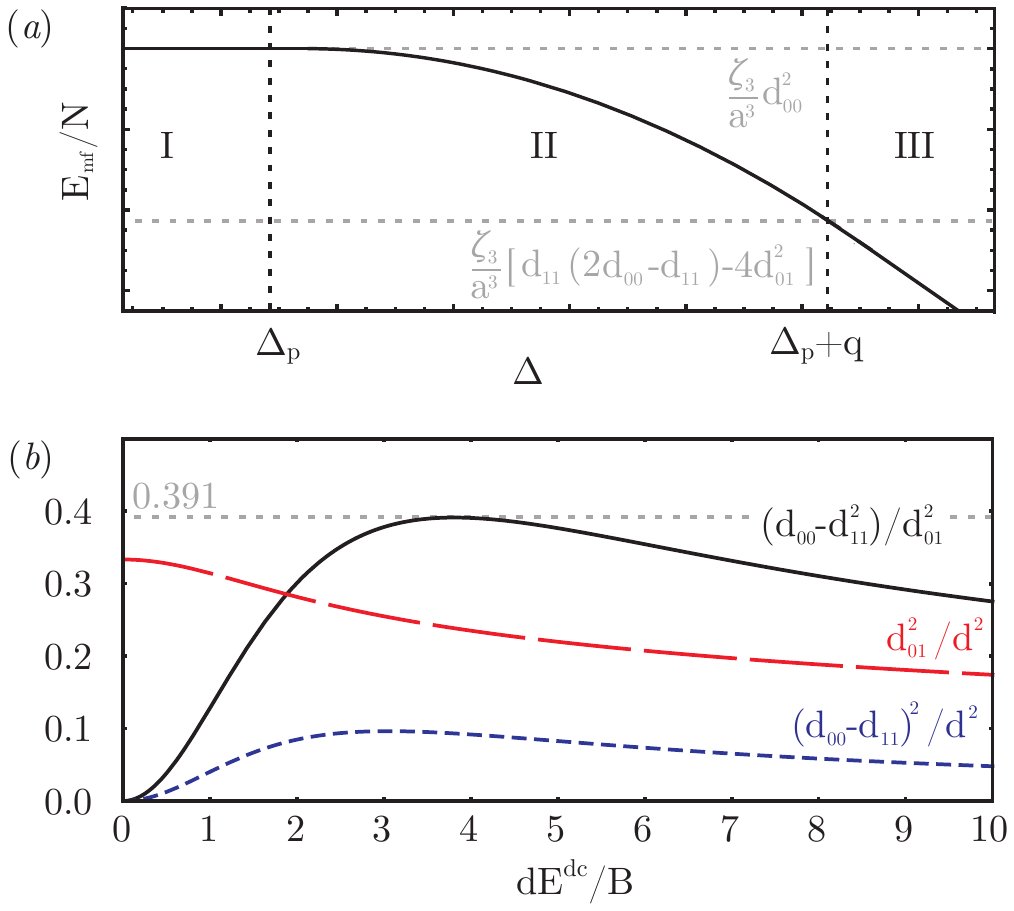}
    \caption{
        \label{fig2}
        (a)~Ground state energy per lattice site in the mean field ansatz showing two continuous quantum phase transitions at $\Delta=\Delta_p$ and $\Delta=\Delta_p+q$.
        (b)~Magnitude of dipole matrix elements, as appearing in the Hamiltonian~\eqref{eq:bosehubbard}, in units of the permanent dipole moment $d$.
        The red line corresponds to the hopping rate $d_{01}^2$, while the blue line
        corresponds to the interaction $(d_{00}-d_{11})^2$.
        The ratio of the two (black) enters the Luttinger parameter as a perturbation term.
    }
\end{figure}

First, we analyze the phase diagram within mean field theory. The general mean-field ansatz takes the form
%
% \begin{eqnarray}
%     |\psi\rangle = \prod_i &\big[\cos(\epsilon)|0\rangle_i + \nonumber\\
%                            &\quad+e^{i\delta}\sin(\epsilon)\left( \cos \theta|{-1}\rangle_i - \sin \theta|1\rangle_i\right)\big], \label{eq:meanfield1}
% \end{eqnarray}
%
\begin{eqnarray}
    |\psi\rangle = \prod_i\! \big[\!\cos \epsilon|0\rangle_i + e^{i\delta}\sin\epsilon( \cos \theta|{-1}\rangle_i - \sin \theta|1\rangle_i)\big] \label{eq:meanfield1}
\end{eqnarray}
with the variational parameters $\epsilon$, $\delta$, and $ \theta$. The ground state is then obtained by
determining the minimum of the energy $E(\epsilon, \delta, \theta) = \langle \psi | H | \psi \rangle$.
It is a special property of the one dimensional geometry, that this variational energy is always minimized for   $\theta = \pi /4$, i.e., the system gives rise to an optimal excitation
\begin{eqnarray}
    |e\rangle =\frac{1}{\sqrt{2}}\left(|1\rangle - |{-1}\rangle \right).
\end{eqnarray}
Furthermore, the $U(1)$ symmetry of the Hamiltonian ensures that the variational energy is independent on the phase $\delta$.
These observations allow us to minimize the energy with respect to the only remaining variational parameter $\epsilon$, and
obtain the ground state energy
\begin{eqnarray}
\frac{E_{\text{mf}}}{N}=
\cases{
\frac{\zeta_3 d_{00}^2}{a^3} & $\Delta<\Delta_p$ \hspace{2.08cm}(I) \\
\frac{\zeta_3 d_{00}^2}{a^3} -\frac{\left(\Delta-\Delta_p\right)^2}{2q} & $\Delta_p<\Delta<\Delta_p+q$ \hspace{0.3cm}(II) \\
\frac{\zeta_3 d_{11}^2}{a^3}-\Delta & $\Delta_p + q<\Delta$ \hspace{1.37cm}(III)
}
\end{eqnarray}
with $\zeta_3=\sum_{j=1}^{\infty}j^{-3}$, see FIG.~\ref{fig2}\,(a). We find two continuous quantum phase transitions  at $ \Delta = \Delta_{p}$
and $\Delta = \Delta_{p}+ q$. The critical values are determined by
\begin{eqnarray}
\Delta_p =-\frac{2\zeta_3}{a^3}\left[d_{00}\left(d_{00}-d_{11}\right)+2d_{01}^2\right], \label{eq:Deltap}
\end{eqnarray}
and the width of the intermediate phase
\begin{eqnarray}
    q = \frac{2\zeta_3}{a^3}\left[\left(d_{00}-d_{11}\right)^2+4d_{01}^2\right]. \label{eq:q}
\end{eqnarray}
This mean field theory confirms the above expectations: for large negative $\Delta$, all molecules are in the ground state with $\epsilon = 0$, which corresponds to phase (I). For increasing detuning $\Delta$, we obtain a phase transition into a phase (II) with a coherent superposition of molecules in the ground state $|0\rangle$ and the optimal excitation $|e\rangle$. This phase breaks the $U(1)$ symmetry and acquires a fixed phase $\delta$ for the superpositions within mean-field. Eventually, a second phase transition appears into phase (III) with all polar molecules excited to the state $|e\rangle$.

The physical interpretation of the intermediate phase is obtained by a study of the dipole operator. It turns out that the latter acquires a macroscopic dipole moment $\mathbf{d}_{xy} = d_{01} \sin(2\epsilon) \cos(\omega t - \delta) \,\mathbf{e}_x$ along the axis of the one-dimensional system, and therefore corresponds to a ferroelectric phase. Note that this dipole moment is oscillating with the frequency $\omega$ of the micro-wave field.

Next we analyze the system beyond mean field theory and analyze the stability of the system towards the formation of crystalline states, where the excitations are ordered periodically in the lattice~\cite{Weimer2010b,Bak1982}. This approach is achieved via a mapping of the Hamiltonian in \eqref{eq:Htotal} onto a Bose-Hubbard model. The basic principle is based on the observation of an optimal excitation $|e\rangle$. We introduce the operators $b^{\dag}_{i}$ creating an optimal excitation at lattice site $i$.  These operators satisfy the well-known hard-core constraint, as only one excitation on each lattice site is allowed. Then, the corresponding Bose-Hubbard model takes the form
\begin{eqnarray}
H=\sum_{ i< j} U_{ij} n_{i}n_j&-\mu \sum_i n_i -\sum_{ i< j}t_{ij}\big(b_i^{\dag}b^{\vphantom\dag}_j+b_j^{\dag}b^{\vphantom\dag}_i\big),  \label{eq:bosehubbard}
\end{eqnarray}
with the number operator $n_{i} = b^{\dag}_{i} b^{\vphantom\dag}_{i}$. Within this notation, the chemical potential takes the form
\begin{eqnarray}
\mu=\frac{2\zeta_3}{a^3}d_{00}(d_{00}-d_{11})+\Delta,
\end{eqnarray}
while the static dipole-dipole interaction provides the Hubbard interaction
\begin{eqnarray}
U_{ij} =\frac{(d_{00}-d_{11})^2}{r^3_{ij}}\label{eq:twopa}
\end{eqnarray}
and the resonant dipolar exchange interactions give rise to an effective hopping
for the optimal excitations with hopping strength
\begin{eqnarray}
    t_{ij}=\frac{2d_{01}^2}{r^3_{ij}}\label{eq:hop} .
\end{eqnarray}
Finally, it is convenient to map the system onto fermions via a Jordan-Wigner-transformation \cite{Jordan1928}.
Introducing the fermionic operators $c^{\dag}_{i}$, the Hamiltonian reduces to
\begin{eqnarray}
H=&\sum_{i< j}U_{ij} n_{i} {n}_j-\mu\sum_i {n}_i
-\sum_{ i< j}t_{ij}\prod_{l=i+1}^{j-1}\!\big(1-2{n}_l\big)\big({c}_i^{\dag}{c}^{\vphantom\dag}_j+{c}_j^{\dag}{c}^{\vphantom\dag}_i\big).  \label{eq:fermionhubbard}
\end{eqnarray}
The Jordan-Wigner transformation produces an additional factor $\prod_{l=i+1}^{j-1}(1-2{n}_l)$ in the hopping term. This factor gives a minus sign for every particle between the lattice sites $i$ and $j$. This term can be separated into a quadratic hopping and a many-body interaction term corresponding to a correlated hopping. Close to the phase transition, the number of excitations is very low, and therefore, we can start with analyzing the quadratic terms in the Hamiltonian and then study the influence of the interaction within perturbation theory. Then, the ground state for the non-interacting fermions is determined by a Fermi sea for the momentum states $\epsilon_{k} < \mu$ with the dispersion relation $\epsilon_{k} = - 4 d_{01}^2 \sum_{i> 0} \cos(r_{i} k) / |\mathbf{r}_{i}|^3$. Note that the dispersion relation exhibits a logarithmic singularity $\epsilon_{k} \sim k^2 \log k$ at small momenta due to the
slow decay of the dipolar interaction.

\begin{figure}[t]
    \centering
    \includegraphics[width=0.6\textwidth]{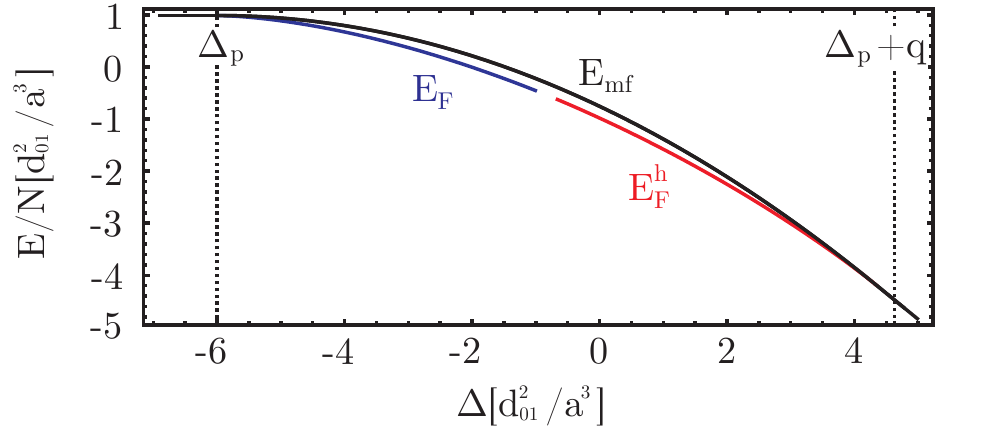}
    \caption{Energy of the mean-field ansatz (black) compared to the energy of the perturbed Fermi sea. The blue line is calculated for a low density of excitations while the red line corresponds to a low density of holes.}
    \label{fig3}
\end{figure}

Including the interaction within first order perturbation theory allows us to determine the ground state energy  exactly up to the corrections of order $n^2$ with $n$ the fermionic density. The resulting ground state energy is shown in  FIG.~\ref{fig3}, and demonstrates that the energy of this perturbative treatment is indeed lower than the variational mean field energy. However,  it is important to stress that this exact treatment reproduces the
mean-field transition point. Finally, we can also reproduce the second phase transition by a particle-hole transformation, i.e., the second phase transition is characterized by a low density of holes, and allows again to determine the energy of the system for small hole densities.

%We computed the energy of the Fermi sea and incorporated the two-particle-interaction and the Jordan-Wigner correction to second order in $n_j$ by first order perturbation theory. To cover both ends of the transition range, we exploited a symmetry between holes and particles in the mapping to the Bose-Hubbard model. The resulting energy is drawn in FIG. \ref{fig:ESFMF}. It reproduced the transition point from zero to one excitation of the mean-field discussion, proving the mean-field ansatz as a good approach in the dilute case. However, an expansion around this point in terms of the number of particles was not possible due to vanishing linear and quadratic terms and a logarithmic divergence in the next higher order. This characteristic divergence was attributed to the $R_{ij}^{-3}$-dependance of the dipole-dipole-interaction.

The mapping onto the Hubbard model provides additional insights into the properties of the ferroelectric phase: first, it allows us to derive the low energy excitations, which are well described by a Luttinger liquid theory. Consequently, the correlation functions decay algebraically and only give rise to quasi long range order as expected  from the Mermin-Wagner theorem for a phase transition with a continuous symmetry in one-dimension. The decay of the ferroelectric correlation function is determined by the Luttinger parameter $K$ via  $\langle b^{\vphantom\dagger}_{i} b^{\dag}_{j} \rangle \sim |{\bf r_{ij}}|^{- K/2}$. In the dilute regime for particles as well as holes, the Luttinger parameter reduces to $K = 1$, while in the intermediate regime it is further reduced by the repulsive
interaction and the corrections from the Jordan-Wigner transformation. We can estimate the Luttinger parameter at half filling using the well established bosonization method \cite{Fradkin2013} to a Hamiltonian including a next-nearest-neighbor hopping correction from the Jordan-Wigner term and a nearest-neighbor interaction,
\begin{eqnarray}
 K^2 \approx  1 + \frac{1}{4\pi} - \frac{(d_{00}-d_{11})^2}{d_{01}^2}\frac{1}{\pi}
\end{eqnarray}
Note that the Luttinger parameter depends on the dipole moments
which are determined by the diagonalization of $H_{\text{rot}}$ in \eqref{eq:Hrot} and
are constraint by the physical implementation of the system.
The different contributions are shown in FIG.~\ref{fig2}\,(b). The fraction determining the ratio of static dipole-dipole-interaction and hopping is limited by $(d_{00}-d_{11})^2/d_{01}^2\lesssim 0.391$. The  minimal achievable value for the Luttinger parameter is therefore bounded by $K \gtrsim 0.977$.
This estimation demonstrates the absence of an additional phase transition within the ferroelectric phase towards a crystalline ordering as the
latter requires values of $K < 1/4$. In other words, in the presented setup, the dipolar exchange interactions (providing the hopping) are stronger than the repulsive dipole-dipole interactions. Consequently, the system prefers a ferroelectric phase with algebraic correlations instead of a crystalline arrangement of the excitations.

\ack
We acknowledge the support from the Deutsche For\-schungs\-gemeinschaft (DFG) within the SFB/TRR~21.

\section*{References}
\providecommand{\newblock}{}

\end{document}